\documentclass[sigconf, nonacm]{acmart}

\usepackage[hyphenbreaks]{breakurl}
\usepackage{amsmath}
\usepackage{amsfonts}
\usepackage{subfigure}
\usepackage{algorithm}
\usepackage{algorithmicx}
\usepackage{algpseudocode}
\usepackage{enumitem}
\usepackage{xspace}

\usepackage{array}
\usepackage{multirow}

\newcommand{\sys}{{\sc Sparcle}\xspace}

\begin{document}
\title{\sys: Boosting the Accuracy of Data Cleaning Systems through Spatial Awareness}

\author{Yuchuan Huang}
\email{huan1531@umn.edu}
\affiliation{
  \institution{University of Minnesota, USA}
  \country{~~~~}
}
\author{Mohamed F. Mokbel}
\email{mokbel@umn.edu}
\affiliation{
  \institution{University of Minnesota, USA}
  \country{~~~~}
}

\sloppy

\begin{abstract}

Though data cleaning systems have earned great success and wide spread in both academia and industry, they fall short when trying to clean spatial data. The main reason is that state-of-the-art data cleaning systems mainly rely on functional dependency rules where there is sufficient co-occurrence of value pairs to learn that a certain value of an attribute leads to a corresponding value of another attribute. However, for spatial attributes that represent locations on the form of {\em <latitude, longitude>}, there is very little chance that two records would have the same exact coordinates, and hence co-occurrence would unlikely to exist. This paper presents \sys~(\underline{SP}atially-\underline{A}wa\underline{R}e \underline{CLE}aning); a novel framework that injects spatial awareness into the core engine of rule-based data cleaning systems as a means of boosting their accuracy. \sys injects two main spatial concepts into the core engine of data cleaning systems: (1)~{\em Spatial Neighborhood}, where co-occurrence is relaxed to be within a certain spatial proximity rather than same exact value, and (2)~{\em Distance Weighting}, where records are given different weights of whether they satisfy a dependency rule, based on their relative distance. Experimental results using a real deployment of \sys inside a state-of-the-art data cleaning system, and real and synthetic datasets, show that \sys significantly boosts the accuracy of data cleaning systems when dealing with spatial data.

\end{abstract}

\maketitle

\begin{figure}[t]
    \centering
    \subfigure[Part of Table of NYC Motor Vehicle Collision Data]{
        \label{fig:NYCExampleTable}
        \includegraphics[width=0.7\linewidth]{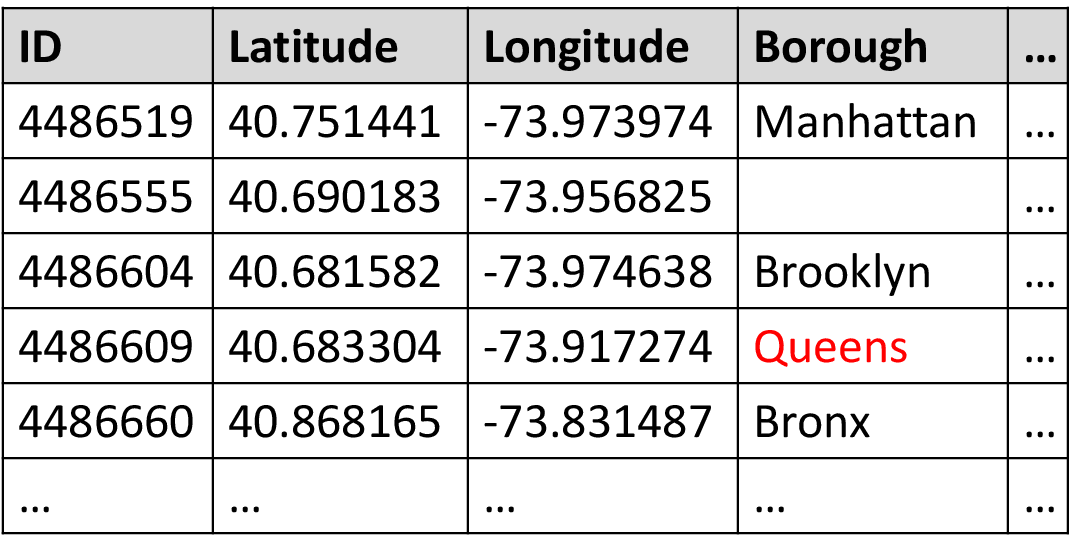}
    }
    \vspace{-5pt}
    \subfigure[Map of NYC Motor Vehicle Collision Data]{
        \label{fig:NYCExampleMap}
        \includegraphics[width=\linewidth,height=0.75\linewidth]{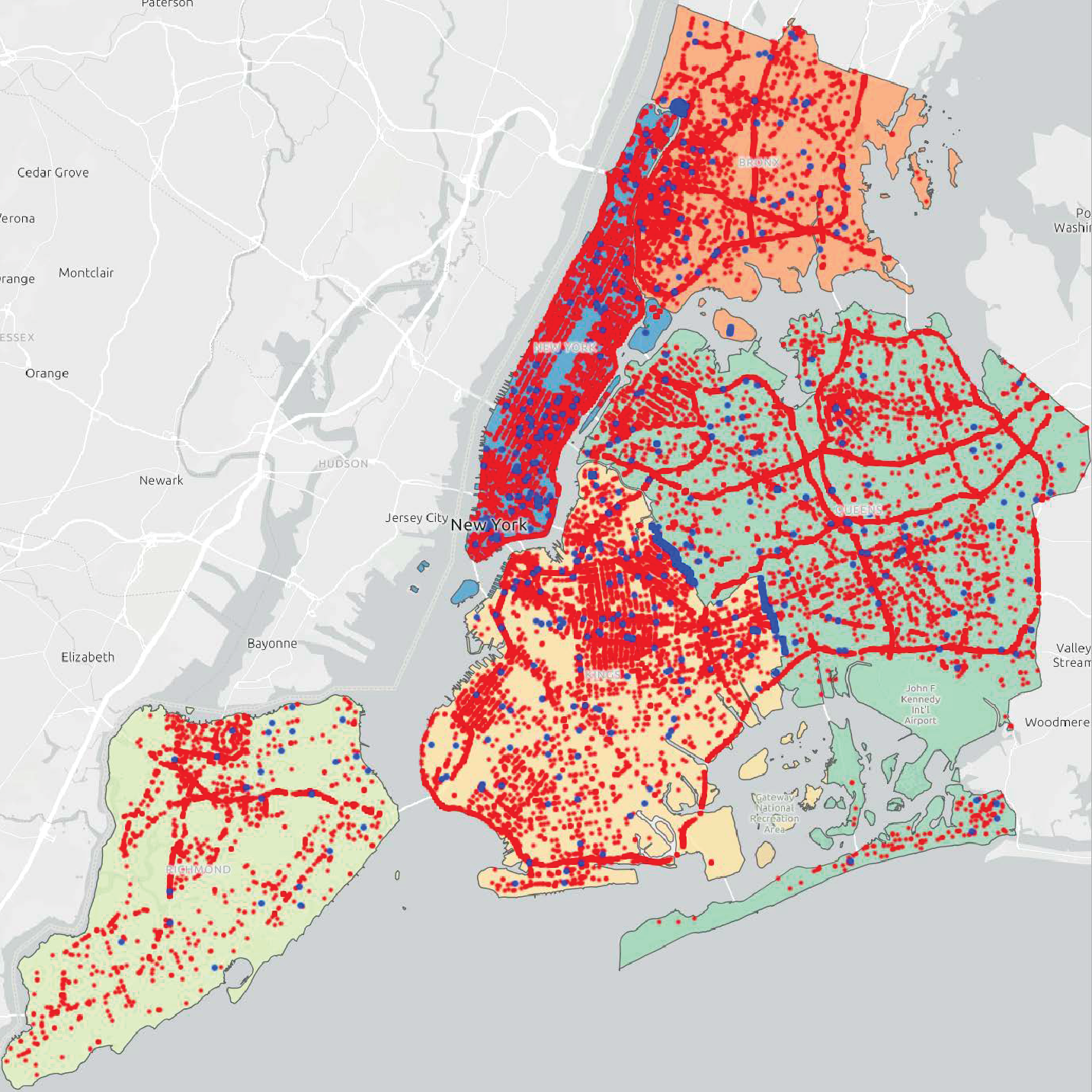}
    }

    \vspace{-10pt}
    \caption{NYC Motor Vehicle Collision Data}
    \vspace{-15pt}
    \label{fig:NYCExample}
\end{figure}

\section{Introduction}
\label{sec:introduction}

Motivated by the inaccuracy and imperfection of real data sets, along with the huge efforts carried by data scientists to manually clean their data, efforts have been dedicated to develop various approaches and systems for automated data cleaning. The goal of such approaches and systems is to boost the quality of real datasets through detecting and repairing erroneous, incomplete, and inconsistent entries. The large majority of such approaches (e.g., see~\cite{BFF+05,CIP13,CM11,GKA20,LKR20,PSC+15}) and systems (e.g., see~\cite{DEE+13,GGM+22,GMP+20,KIJ+15,MA20,RCI+17,ROA+21}) are rule-based, where functional dependencies between various attributes are defined as rules to guide the data cleaning process. The success and immense need of such data cleaning approaches and systems have made it widely adopted by industry~\cite{MS, GCP, AWS, IBM} and commercial startups~\cite{Inductiv,Tamr,Trifacta}. 

Unfortunately, with all its success and wide spread, state-of-the-art data cleaning systems and approaches fall short when trying to clean data with spatial attributes dependency. As an example, we investigated the NYC Motor Vehicle Collision data~\cite{NYCOpenData}, which includes 1,751,624 collision records that took place in the New York City since 2014. A real snapshot of this dataset is shown in Figure~\ref{fig:NYCExampleTable} for five collision records and only four attributes of each collision ({\em ID}, {\em Latitude}, {\em Longitude}, {\em Borough}). The snapshot shows two kinds of errors: (1)~the second record is missing the {\em Borough} information, and (2)~the fourth record has the wrong {\em Borough} information. To get an idea of the scale of the problem, Figure~\ref{fig:NYCExampleMap} plots all the erroneous collision records over NYC map (421,013 records, i.e., 24$\%$ of the data), where 418,896 records have a missing borough (plotted in red) and 2,117 records have incorrect borough information (plotted in blue). We fed this data along with the functional dependency:  ({\em Latitude}, {\em Longitude}) $\rightarrow$ {\em Borough} to HoloClean~\cite{RCI+17} system as a state-of-the-art rule-based data cleaning framework. HoloClean was able to only correct 58.7$\%$ of the errors, which is a pretty low accuracy compared to the ability of HoloClean in cleaning non-spatial data with more than 95$\%$ accuracy~\cite{RCI+17}. To understand such poor accuracy of HoloClean when dealing with spatial data, we distinguish between: (a)~those erroneous records that took place in the {\em same exact} location of at least one other correct collision record, and (b)~those erroneous records that took place in {\em new} locations where there is no other correct locations records. As depicted in the second column of Table~\ref{tbl:NYCResult}, HoloClean was able to correct 99.6$\%$ of the former, but only 30.3$\%$ of the latter. 

\begin{table}
    \centering
    \begin{tabular}{l|c|c}
        \toprule
                                                   & HoloClean & \sys            \\
        \midrule
        \textbf{Total}                             & 58.7\%    & \textbf{99.4\%} \\
        \hspace{8pt} Errors at duplicated location & 99.6\%    & \textbf{99.7\%} \\
        \hspace{8pt} Errors at new location        & 30.3\%    & \textbf{99.1\%} \\
        \bottomrule
    \end{tabular}
    \caption{Error Repairing of NYC Motor Vehicle Collision Data}
    \vspace{-15pt}
    \label{tbl:NYCResult}
\end{table}

The main reason behind such poor performance of HoloClean, as a representative of rule-based data cleaning systems, when dealing with spatial data is twofold: (1)~Cleaning with functional dependencies relies on sufficient co-occurrence of value pairs to learn that a certain value of an attribute leads to a corresponding value of another attribute. However, for spatial attributes, there is very little chance that two records have the exact same coordinates, mainly due to the inherent inaccuracy of the devices generating the locations. Hence, a rule-based system will not be able to find sufficient {\em spatial} co-occurrence to be used to detect and repair erroneous entries. (2)~In rule-based systems, the outcome of whether a certain record satisfies a rule is binary ({\em True} or {\em False}). However, in spatial rules, such outcome needs to be fuzzy, as a certain record may satisfy the rule in stronger terms than other records. It is important to note that in this particular example, we do not rely on any external knowledge of borough boundaries for a couple of practical reasons. First, rule-based data cleaning systems usually run as an automated process based on its data input with no additional procedure that rely on external knowledge, and our goal is to keep it working as is. Second, for many applications, the boundary information is unknown and needs to be inferred. For example, regions with various levels of air quality, where there will still be a functional dependency between the location and the air level quality level, yet, the boundaries of various air quality regions are unknown.

This paper presents \sys (\underline{SP}atially-\underline{A}wa\underline{R}e \underline{CLE}aning); a novel framework that injects spatial awareness into the core engine of rule-based data cleaning systems as a means of boosting their accuracy. \sys aims to go beyond the traditional functional dependency rules of the form:  {\em ``Two records with the \textbf{same} location \textbf{should} have the same borough''} to support a more relaxed functional dependency form that is more suitable to the nature of spatial data: {\em ``Two records with \textbf{more similar} locations are \textbf{more likely} to have the same borough''}. To do so, \sys injects the following two main spatial concepts into the core engine of data cleaning systems: (1)~{\em Spatial Neighborhood}. To support going from the {\em ``same''} predicate in traditional functional dependency to the {\em ``similar''} predicate in the relaxed functional dependency, records with spatial attributes satisfying some spatial neighborhood (similarity) criteria (e.g., within one mile of each other) should be considered as relatively equivalent with respect to the spatial functional dependencies; (2)~{\em Distance Weighting}. To support going from {\em ``should''} to {\em ``likely''} and to have the keyword {\em ``more''} in the relaxed functional dependency, records will be given a weight of how much they satisfy each rule, where the weight will be based on the distance between records satisfying the functional dependency. With both the {\em spatial neighborhood} and {\em distance weighting} concepts, the last column of Table~\ref{tbl:NYCResult} shows that, for the NYC vehicle collision data, \sys was able to correct $99.4\%$ of all errors, and $99.1\%$ of the errors with new locations (compare this to $58.7\%$ and $30.3\%$ of HoloClean).

As \sys aims to inject spatial awareness into existing data cleaning systems, its architecture follows the same common architecture of most rule-based data cleaning systems. In particular, such systems (e.g.,~\cite{DEE+13,GGM+22,KIJ+15,MA20,RCI+17,YBE13}) are typically composed of four back-to-back components, {\em error detector}, {\em candidate generator}, {\em input formulator}, and {\em error corrector}. The first three modules are mainly for error detection and preparing the data in some format that can be repaired using a statistical method in the last module. Hence, internally, \sys focuses on the modifying the first three modules to be: {\em spatial error detector}, {\em spatial candidate generator}, and {\em spatial signal formulator}. The {\em spatial error detector} module is responsible on defining the parameters of both the spatial neighborhood and distance weighting concepts, as well as taking spatial neighborhood into account when detecting errors in the input dataset. The {\em spatial candidate generator} module exploits both concepts to generate a set of candidate values for each erroneous entry detected from the spatial error detector module. The {\em spatial input formulator} module goes through each candidate value and uses both the spatial neighborhood and distance weighting concepts to give appropriate score for each value that reflects how likely it is to be the correct value according to the spatial dependency rule. The output of this module is passed to the underlying data repair module that would statistically decide on the correct value based on the scores given from \sys spatial input formulator module along with other non-spatial dependency rules.

Extensive experiments based on real implementation of \sys inside a data cleaning system~\cite{RCI+17} and real datasets from Austin, Chicago, and New York City show that \sys significantly boosts the accuracy of state-of-the-art data cleaning systems, namely HoloClean~\cite{RCI+17} and Baran~\cite{MA20}. The rest of the paper is organized as follows: Section~\ref{sec:architecture} presents \sys architecture. The three modules of \sys are presented in Sections~\ref{sec:ErrorDetector} to~\ref{sec:InputFormulator}. Experimental results are presented in Section~\ref{sec:Experiment}. Section~\ref{sec:relatedWork} highlights related works to \sys. Finally, Section~\ref{sec:conclusion} concludes the paper.



\section{Architecture}
\label{sec:architecture}

Figure~\ref{fig:architecture} depicts the \sys architecture. \sys takes two types of inputs, the raw data that need to be cleaned, and the constraints that define the functional dependencies. As \sys is injected into a host rule-based data cleaning system (e.g.,~\cite{GGM+22,MA20,RCI+17,WZI+20}), it only takes care of the spatial constraints, while the non-spatial constraints are supported by the host data cleaning system. The output of \sys is basically the input data set, yet, with detected erroneous {\em cells}, where a {\em cell} is a certain attribute of a certain record, along with a list of suggested correct values for each cell. Each suggested value has a score that indicates how likely that value is the correct one for its corresponding cell. This means that if we have only spatial constraints, then the output of \sys would be the completely corrected input data. However, as data is likely to have other non-spatial constraints, the output of \sys is sent to the {\em error correction} module of its host data cleaning system. Then, it will be integrated with other suggested values from the non-spatial constraints to statistically come up with the final correct value. Internally, \sys follows similar architecture to that of rule-based data cleaning systems, that is mostly composed of three main modules, {\em error detector}, {\em candidate generator}, and {\em input formulator}. Yet, \sys injects spatial-awareness into the logic and core execution of each of these modules. Hence, \sys architecture (Figure~\ref{fig:architecture}) is mainly composed of the following three modules:

\begin{figure}[t]
    \centering
    \includegraphics[width=\linewidth]{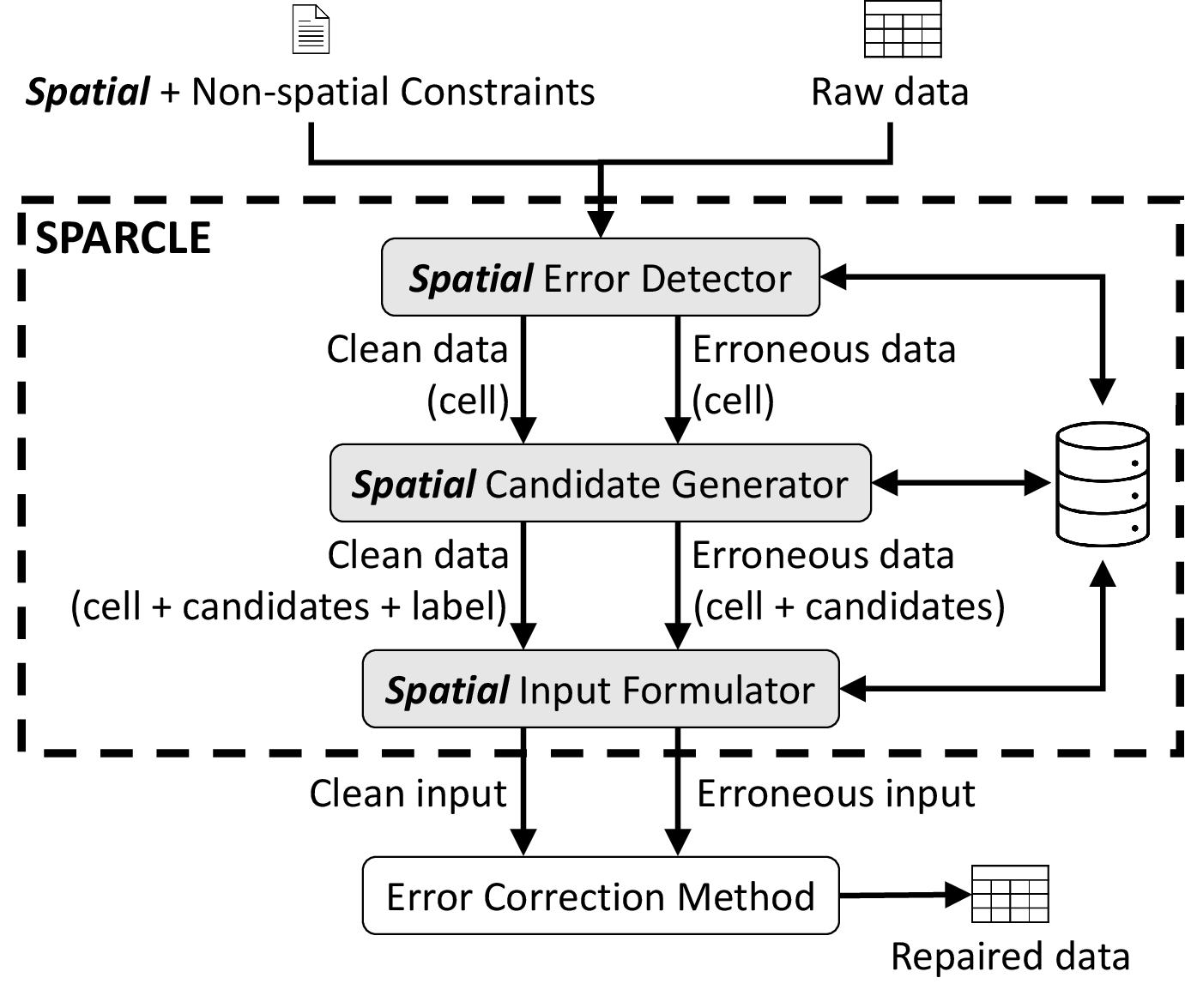}
    \vspace{-25pt}
    \caption{\sys Architecture}
    \vspace{-25pt}
    \label{fig:architecture}
\end{figure}

\vspace{2pt}
\noindent \textbf{Spatial Error Detector.} The input to this module is the input to \sys. The output is two sets of cells, erroneous and clean cells. It injects spatial-awareness into existing error detection modules. Hence, instead of detecting errors based on exact co-occurrence, it relaxes the co-occurrence criteria to consider records within spatial proximity. It also assigns weights to all detected errors based on the distance between co-occurred records. Details are in Section~\ref{sec:ErrorDetector}.

\vspace{2pt}
\noindent \textbf{Spatial Candidate Generator.} The input to this module is the two sets of erroneous and clean cells coming out of the {\em spatial error detector} module. The output is two similar sets of cells, considering the following: (1)~There will be more cells in the clean set, as some erroneous cells will be cleaned, (2)~Each cell will have a set of candidate values, where \sys believes that one of these values in the correct value, (3)~Clean cells will be labeled with the value that \sys believes it is the correct one. The module injects spatial-awareness into existing candidate generation modules. So, instead of getting candidates from records with same locations, candidates will be drawn from records within spatial neighborhood. Each candidate value is given a weighted probability based on its distance from the erroneous cell. Details are in Section~\ref{sec:CandidateGenerator}.

\vspace{2pt}
\noindent \textbf{Spatial Input Formulator.} The input to this module is the output of the {\em spatial candidate generator} module, while the output is the output of \sys. The module mainly injects spatial-awareness into existing input formulator modules. As these modules are very specific to the host data cleaning systems, \sys has to have various versions of such module to match its host system. The goal of the {\em input formulator} is to score each possible candidate value and prepare the output in a certain format to match the requirements of the error correction module. \sys is equipped with input formulators for three categories of data correction modules, namely, violation-based feature vectors (used in AimNet data cleaning system~\cite{WZI+20}), probability-based feature vectors (used in Baran data cleaning system~\cite{MA20}), and factor graphs (used in HoloClean~\cite{RCI+17} and MLNClean~\cite{GGM+22} data cleaning system). Details are in Section~\ref{sec:InputFormulator}.

\section{Spatial Error Detector}
\label{sec:ErrorDetector}

The {\em spatial error detector} module is the first module in \sys, where its input is the input to \sys, which is the dataset to be cleaned, along with a set of spatial constraints that need to be enforced for error detection and correction. The output are two sets of {\em cells}, where a {\em cell} is basically a certain attribute of a certain record. The first set of {\em cells} is the set of input {\em cells} that are deemed erroneous, while the second set of {\em cells} are all other {\em cells} that are considered clean for now. The module enriches existing error detection modules in state-of-the-art data cleaning systems~\cite{CIP13,RCI+17,WZI+20,GGM+22} by: (1)~providing the ability to incorporate both the {\em Spatial Neighborhood} and {\em Distance Weighting} concepts, along with their parameters, in the language defining the spatial constraints (Section~\ref{subsec:Constraints}), (2)~building spatial data infrastructure that boosts the performance and scalability of \sys to support spatial constraints (Section~\ref{subsec:DataStructure}), and (3)~employing the spatial neighborhood concept to detect erroneous {\em cells} that violate the spatial constraints and the distance weighting concepts to assign a weight to each constraint violation (Section~\ref{subsec:Detection}).

\subsection{Spatial Denial Constraints}
\label{subsec:Constraints}

\noindent\textbf{Denial Constraints.} Most rule-based data cleaning systems (e.g.,~\cite{CIP13,RCI+17,GGM+22}) use denial constraints as their way of expressing the cases that should be denied for any dataset instance. A denial constraint of the form $\forall r_1, r_2, \dots \in \mathcal{R}: \neg(p_1 \land p_2 \land \dots \land p_m)$ means that the set of predicates $p_1$ to $p_m$ cannot be {\em True} together for any combination of records in the dataset, where the result of a predicate evaluation is either {\em True} or {\em False}. For the example of NYC collision data given in Figure~\ref{fig:NYCExampleTable}, a denial constraint in typical data cleaning systems would be:

\vspace{-6pt}
\begin{displaymath}
    \begin{aligned}
        \forall r_1, r_2 \in R: \neg (
         & r_1.Latitude = r_2.Latitude \:\land   \\
         & r_1.Longitude = r_2.Longitude \:\land \\
         & r_1.Borough \neq r_2.Borough)
    \end{aligned}
\end{displaymath}
\vspace{-6pt}

, which indicates that two records $r_1$ and $r_2$ cannot have the same values for both latitude and longitude, but have different borough values. In other words, if $r_1$ and $r_2$ have the same latitude and longitude values, they should have the same borough value.

\vspace{2pt}
\noindent\textbf{Spatial Denial Constraints.} Unfortunately, the current form of denial constraints is not helpful for spatial data, where it is rare that two records would have the same exact latitude and longitude values. As \sys aims to provide native spatial data support, it extends the denial constraint language to support spatial denial constraints. For the NYC collision data in Figure~\ref{fig:NYCExampleTable}, \sys supports the following construct for a {\em SpatialRange} denial constraint:

\vspace{-6pt}
\begin{displaymath}
    \begin{aligned}
        \forall r_1, r_2  \in R: \neg( & SpatialRange(r_1.Latitude, r_1.Longitude,                                                \\
                                       & \qquad \qquad \qquad r_2.Latitude, r_2.Longitude, d, {\mathcal F}, {\mathcal W}) \:\land \\
                                       & r_1.Borough \neq r_2.Borough)
    \end{aligned}
\end{displaymath}
\vspace{-6pt}

, which indicates that if two records $r_1$ and $r_2$ are within a certain distance range $d$ from each other, according to distance function ${\mathcal F}$, then they are {\em likely}, according to weight function ${\mathcal W}$, to have the same borough. The distance function ${\mathcal F}$, responsible on enforcing the {\em spatial neighborhood} concept, may employ either Euclidean or road network distance. The weight function ${\mathcal W}(r_1, r_2)$, responsible on enforcing the {\em distance weighting} concept, by employing an arbitrary function (e.g., linear or exponential) that returns a decreasing value from 1 to 0 as the distance between $r_1$ and $r_2$ increases from 0 to $d$. {\em SpatialRange} returns {\em True} as a non-zero value (based on ${\mathcal W}$) if ${\mathcal F}(r_1, r_2) < d$, otherwise it returns {\em False}.

\sys also supports $k$-{\em nearest-neighbor} ($k$NN) denial constrains with the following language construct:

\vspace{-6pt}
\begin{displaymath}
    \begin{aligned}
        \forall r_1, r_2 \in R: \neg ( & SpatialkNN(r_1.Latitude, r_1.Longitude,                                                 \\
                                       & \qquad \qquad \quad r_2.Latitude, r_2.Longitude, k, {\mathcal F}, {\mathcal W}) \:\land \\
                                       & r_1.Borough \neq r_2.Borough)
    \end{aligned}
\end{displaymath}
\vspace{-6pt}

, which indicates that if record $r_2$ is among the $k$ nearest records to $r_1$, according to distance function ${\mathcal F}$, then they are {\em likely}, according to weight function ${\mathcal W}$, to have the same borough. {\em SpatialkNN} returns {\em True} as a non-zero value (based on ${\mathcal W}$) if $r_2$ is among the $k$ nearest records to $r_1$, otherwise it returns {\em False}.

\subsection{Building Data Infrastructure}
\label{subsec:DataStructure}

Current data cleaning systems check their constraint violations by simply doing a self-join for the input dataset based on equality for one of the predicates, then, scanning the result for other predicates. Further operations in the data cleaning process are basically employing inexpensive equality search. Unfortunately, this is not the case for evaluating spatial constraints violation along with downstream operations that will be needed in \sys. In particular, the most needed operations in \sys are spatial range and $k$-nearest-neighbor queries per the underlying spatial constraints. Since these are pretty expensive operations, compared to the equality search, \sys: (a)~employs a spatial database system~\cite{PostGIS}, where the input raw data is spatially indexed based on their {\em Latitude} and {\em Longitude} coordinates, and (b)~for each spatial constraint $C$, \sys uses the spatial index to efficiently perform a self-join of the raw input data based on either range or $k$-nearest-neighbor query with the parameters defined in $C$. The result of the join is then materialized and stored in a table, termed {\em DistanceMatrix}, with the schema: (\underline{$R_1$}, \underline{$R_2$}, $v_1$, $v_2$, $D$, $W$), where $R_1$ and $R_2$ are two record identifiers such that $R_2$ is either within distance $d$ from $R_1$ or is one of the $k$ nearest neighbors of $R_1$, according to a distance function ${\mathcal F}$ as defined by range or $k$-nearest neighbor spatial constraints in Section~\ref{subsec:Constraints}. $v_1$ and $v_2$ are the corresponding values for $R_1$ and $R_2$ for the attribute mentioned in $C$ that we aim to clean. $D$ is the distance between $R_1$ and $R_2$ according to the function ${\mathcal F}$ and $W$ is the weight for the distance between $R_1$ and $R_2$ according to the weight function $\mathcal{W}$.

\begin{figure}[t]
    \centering
    \includegraphics[width=\linewidth]{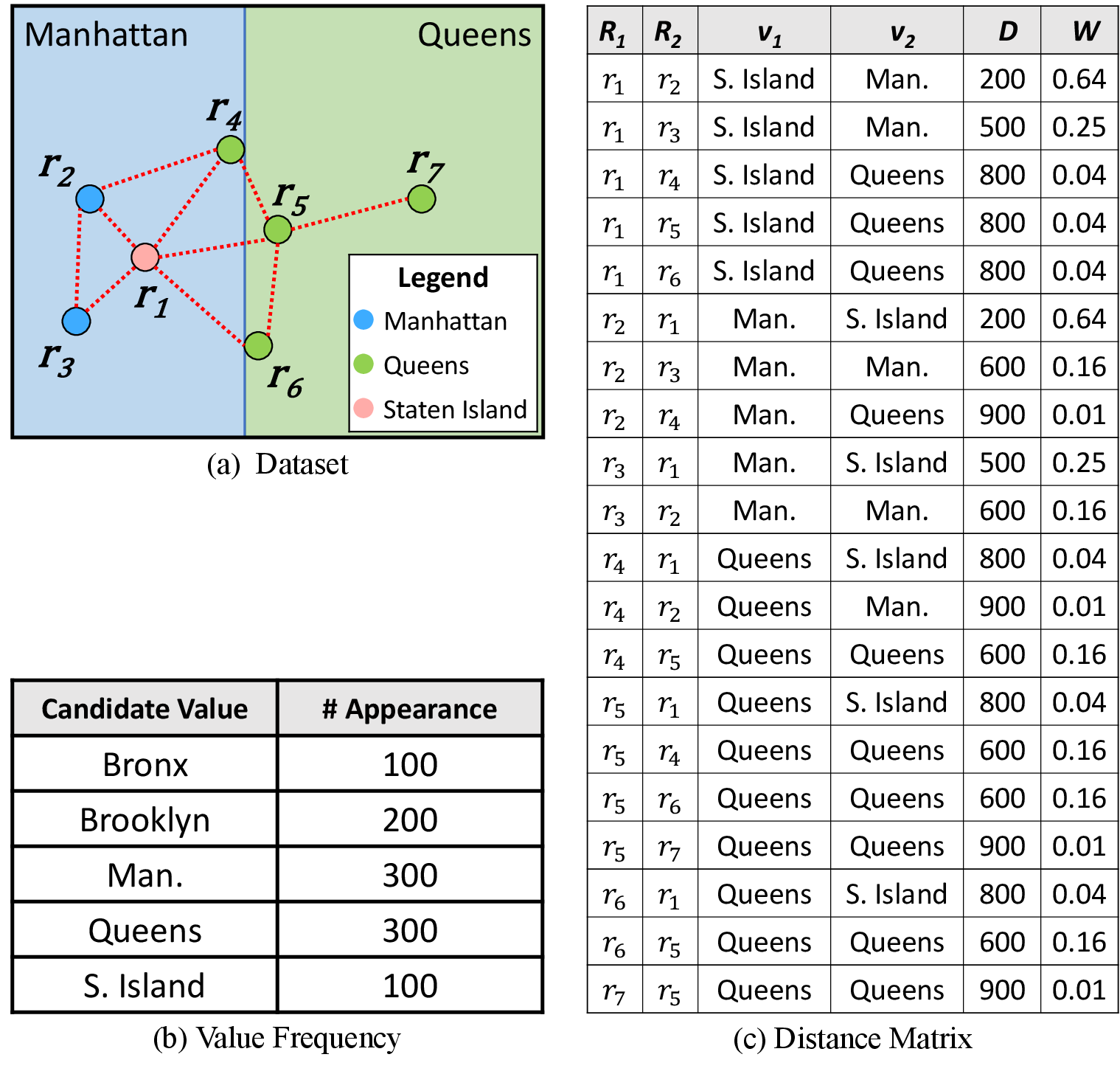}
    \vspace{-17pt}
    \caption{DistanceMatrix Example}
    \vspace{-17pt}
    \label{fig:DMExample}
\end{figure}

Figure~\ref{fig:DMExample} gives an example for the {\em DistanceMatrix} computations. In particular, Figure~\ref{fig:DMExample}a shows a set of seven records on part of the map that includes areas from two NYC boroughs, Manhattan and Queens, plotted in light blue and green, respectively. It is important to note that the borough boundaries are not known to the data cleaning system, and they are just depicted here for illustration of the ground truth, but this information is not used at all in any of \sys computations. Meanwhile, records are colored based on the borough information they have in their raw records, which could be right or wrong. Figure~\ref{fig:DMExample}b gives statistics about the whole dataset, in terms of the number of records in the dataset for each borough value. These statistics, collected in this module, will be used by later modules for their computations. Figure~\ref{fig:DMExample}c gives the {\em DistanceMatrix} for the seven records ($r_1$ to $r_7$) of Figure~\ref{fig:DMExample}a, based on a {\em SpatialRange} denial constraint with distance $d$=1km (plotted as red dashed lines between records), and a weight function $W(r_1, r_2) = (1 - \frac{\mathcal{F}(r_1, r_2)}{d})^2$, where $\mathcal{F}(r_1, r_2)$ is the distance between $R_1$ and $R_2$. For $r_1$, with borough value Staten Island, there are five records ($r_2$ to $r_6$) within distance $d$. Hence, there are five corresponding records in the {\em DistanceMatrix}. Two of these records, $r_2$ and $r_3$ have borough value Manhattan, with distances 200m and 500m, which result in weight values 0.64 and 0.25, respectively. Three of these records ($r_4$ to $r_6$) have distance 800m from $r_1$, and hence they would have weight 0.04. In a similar way, $r_2$, $r_3$, $r_4$, $r_5$, $r_6$, and $r_7$ would have three, two, three, four, two, and one records in the {\em DistanceMatrix}.

\subsection{Detecting Spatial Constraint Violation}
\label{subsec:Detection}

Algorithm~\ref{alg:ErrorDetection} gives the pseudo code of \sys spatial error detector. With the {\em DistanceMatrix}, computed by a self-join as described in Section~\ref{subsec:DataStructure}, the spatial error detection module becomes straightforward and pretty inexpensive. It is basically one scan over the {\em DistanceMatrix}, where for each record ($R_1$, $R_2$, $v_1$, $v_2$, $D$, $W$), if $v_1$ $\neq$ $v_2$, we consider that both $R_1$.{\em Borough} and $R_2$.{\em Borough} are erroneous cells. Hence we move these two cells from the set of all (clean) cells to the set of erroneous ones. Once we finish a full scan, we output both sets of cells. The rationale behind this is that each row in the {\em DistanceMatrix} refers to two records that satisfy the spatial predicate (i.e., within range distance or $k$-nearest neighbor). Hence, they are expected to have the same borough value (i.e., $v_1$ is likely to be the same as $v_2$). If not, then at least one of these two records might have wrong borough value. Since we are not sure which one is wrong, we put both cells in the erroneous set.

For the example in Figure~\ref{fig:DMExample}, all the {\em Borough} cells (attributes) of records $r_1$ to $r_6$ will be added to the set of erroneous cells as they appear in the first five rows of the {\em DistanceMatrix}, where $v_1$ $\neq$ $v_2$. $r_7$.{\em Borough} will be considered a clean cell as all its rows in the {\em DistanceMatrix} have $v_1$ $=$ $v_2$.

\begin{algorithm}[t]
    \begin{flushleft}
        \noindent \textbf{Procedure} SpatialErrorDetection(Data ${\mathcal I}$, DistMatrix $M$)
    \end{flushleft}
    \begin{algorithmic}[1]
        \State {\em ErroneousCells} $\leftarrow$ $\phi$
        \State {\em CleanCells}  $\leftarrow$ All cells in Input Data ${\mathcal I}$
        \For {\textbf{each} row ($R_1$, $R_2$, $v_1$, $v_2$, $D$, $W$) \textbf{in} $M$}
        \If {$v_1$ $\neq$ $v_2$}
        \State Move $R_1$.{\em Borough} from {\em CleanCells} to {\em ErroneousCells}
        \State Move $R_2$.{\em Borough} from {\em CleanCells} to {\em ErroneousCells}
        \EndIf
        \EndFor
        \State \textbf{Return} {\em CleanCells} and {\em ErroneousCells}
    \end{algorithmic}
    \caption{Error Detection Pseudo Code}
    \label{alg:ErrorDetection}
\end{algorithm}

\section{Spatial Candidate Generator}
\label{sec:CandidateGenerator}

Though this module takes its input from the {\em Spatial Error Detector} module (Section~\ref{sec:ErrorDetector}) as two sets of cells, {\em clean} and {\em erroneous}, it mainly operates on the {\em erroneous} list aiming to: (1)~generate a list of candidate values for each erroneous cell, along with the probability that each candidate is the correct one, (2)~use the probabilities of the generated candidates to decide if any of the erroneous cells can be safely moved to the list of clean cells. The output would be another two sets of cells, {\em clean} and {\em erroneous}, along with the generated candidate values for each cell. This is a pretty standard procedure in rule-based data cleaning systems~\cite{GGM+22,MA20,RCI+17,WZI+20}. Yet, standard procedures cannot capture the spatial properties of input data, and hence miss generating important candidate values. Hence, \sys injects the spatial-awareness with its two concepts, {\em spatial neighborhood} and {\em distance weighting} in the candidate generation process. In particular, this process goes through three phases as outlined in Algorithm~\ref{alg:CandGen}. The first phase (Section~\ref{subsec:Generation}) generates an initial list of possible candidate values for each erroneous cell. The second phase (Section~\ref{subsec:Probability}) estimates a probability of correctness for each candidate value. The third phase (Section~\ref{subsec:Labeling}) finds if there is one clear dominant candidate value. If so, it is considered as the correct value of its cell, and the cell is moved to the clean list.

\subsection{Phase 1: Initial Candidate Generation}
\label{subsec:Generation}

Data cleaning systems mainly generate the candidate values based on {\em counting} the {\em co-occurrence} between record attributes. For example, the value of $r_i$.{\em Borough} would be a likely candidate of $r_j$.{\em Borough} if $r_i$ and $r_j$ share the same {\em Latitude} value (i.e., {\em co-occurrence} of latitudes). The likeliness of the candidacy will be based on the {\em counting} of how many times such co-occurrence took place. Apparently, this is not applicable to spatial data as it is rare to have two records with the same {\em Latitude} and/or {\em Longitude} values. In fact, applying this to the example in Figure~\ref{fig:DMExample} yields zero co-occurrence and hence no candidates are generated for any of the erroneous cells.

\sys enriches existing candidate generators with spatial awareness. In particular, for any record $r_i$ where its cell/attribute {\em Borough} is marked erroneous, \sys modifies the candidate generation process in two ways: (1)~The {\em co-occurrence} of record values is relaxed from {\em exact} value co-occurrence to be {\em nearby} co-occurrence. Hence, the candidate values for $r_i$.{\em Borough} would include $r_i$.{\em Borough} itself, along with all {\em Borough} values for any record $r_j$ that lies within (range or $k$NN) proximity from $r_i$ according to the spatial denial constraint. This is done through a lookup search over the {\em DistanceMatrix} for all rows where $R_1$ is $r_i$. (2)~The {\em counting} of the co-occurrence is relaxed from being an {\em absolute} count to be a {\em weighted} count based on how far the co-occurred records from each other. This can be done by computing the sum of the weights in the {\em DistanceMatrix} for all co-occurred records, i.e., all rows where $R_1$ is $r_i$. If none of the nearby records share the same value of $r_i$.{\em Borough}, we would still have the value of $r_i$.{\em Borough} in our candidate list, yet with a default minimal weight value of 0.01. The weights of all candidate values will be used in the second phase (Section~\ref{subsec:Probability}) to estimate the probability of each candidate value.

\begin{algorithm}[t]
    \begin{flushleft}
        \noindent \textbf{ Procedure} SpatialCandGeneration(Cells ${\mathcal C}$, Cells ${\mathcal E}$, {\em MaxProb})
    \end{flushleft}
    \begin{algorithmic}[1]       
        \For {\textbf{each} erroneous cell $E$ in ${\mathcal E}$}            
            \State $E$.{\em CandList} $\leftarrow$  {\em InitCandidates}($E$) \Comment{\textbf{(Phase~1)}}
            \For {$i$=1 to  $\mid$$E$.{\em CandList}$\mid$} \Comment{\textbf{(Phase~2)}}
                \State $E$.{\em CandList[i].Prob} $\leftarrow$ {\em ProbEval}($E$.{\em CandList[i].value})
            \EndFor
            \State ProbNormalization($E$.{\em CandList}) \Comment{\textbf{(Phase~3)}}
            \If {$\mid$$E$.{\em CandList}$\mid$ = 1 \textbf{or} TopProb($E$.{\em Candlist}) > {\em MaxProb}}
                \State $E$.{\em Label} $\leftarrow$ TopProbCandidate($E$.{\em Candlist})
                \State Move $E$ from ${\mathcal E}$ to ${\mathcal C}$
            \EndIf
        \EndFor
        \State \textbf{Return} ${\mathcal E}$ and ${\mathcal C}$
    \end{algorithmic}
    \caption{Candidate Generation Pseudo Code}
    \label{alg:CandGen}
\end{algorithm}

\vspace{1pt}
\noindent\textbf{Example.} The second and third columns of Table~\ref{tbl:CandGen} give the list of candidate values, along with their weights for all the six erroneous cells in Figure~\ref{fig:DMExample}, namely, the {\em Borough} values for $r_1$ to $r_6$. For $r_1$, there are three candidate values for its {\em Borough}, Manhattan and Queens as they appear two and three times, respectively, in the {\em DistanceMatrix} with nearby records (i.e., nearby co-occurrence). The third candidate value is Staten Island, even though no nearby record has this value, but it is the raw {\em Borough} value of $r_1$, and hence we need to consider it. The weights for Manhattan and Queens are set to 0.89 and 0.12, respectively, computed as the sum of weights of their corresponding records in the {\em DistanceMatrix}. The weight for Staten Island is set to the default minimal value of 0.01. For $r_2$, $r_3$, $r_4$, and $r_6$, there are 3, 2, 3, and 2 candidate values, respectively, each appears (co-occurs) once in the {\em DistanceMatrix}, and hence their weights are just copied. For $r_5$, there are two candidates, Queens, which appears three times with $r_5$ in the {\em DistanceMatrix} with a sum of weights 0.33, and Staten Island, which appears only once with $r_5$. Notice that $r_7$ is not included here as it was already marked clean by the {\em spatial error detector} module.

\begin{table*}[]
  \centering
  \begin{tabular}{l|rrrrrr}
    \toprule
    \textbf{Cell}          & \textbf{Candidate Value} & \textbf{Sum Weights} & $\frac{|Spatial(v, R)|}{|D|}$ & $ \frac{Count((v, R.ID), D)}{Count(v, D)}$ & \textbf{Probability} & \textbf{Normalized Prob.} \\
    \midrule
    \multirow{3}{*}{$r_1$} & Manhattan                & 0.89                   & 0.00089                       & 0.1/300                                    & 89/300000000       & 0.68               \\
                           & Queens                   & 0.12                   & 0.00012                       & 0.1/300                                    & 1/25000000       & 0.09               \\
                           & S. Island                & 0.01                   & 0.00001                       & 1/100                                      & 1/10000000         & 0.23               \\
    \midrule
    \multirow{3}{*}{$r_2$} & Manhattan                & 0.16                   & 0.00016                       & 1/300                                      & 1/1875000          & 0.45               \\
                           & Queens                   & 0.01                   & 0.00001                       & 0.1/300                                    & 1/300000000        & 0.01               \\
                           & S. Island                & 0.64                   & 0.00064                       & 0.1/100                                    & 1/1562500          & 0.54               \\
    \midrule
    \multirow{2}{*}{$r_3$} & Manhattan                & 0.16                   & 0.00016                       & 1/300                                      & 1/1875000          & 0.68               \\
                           & S. Island                & 0.25                   & 0.00025                       & 0.1/100                                    & 1/4000000          & 0.32               \\
    \midrule
    \multirow{3}{*}{$r_4$} & Manhattan                & 0.01                   & 0.00001                       & 0.1/300                                    & 1/300000000        & 0.01               \\
                           & Queens                   & 0.16                   & 0.00016                       & 1/300                                      & 1/1875000          & 0.92               \\
                           & S. Island                & 0.04                   & 0.00004                       & 0.1/100                                    & 1/25000000         & 0.07               \\
    \midrule
    \multirow{2}{*}{$r_5$} & Queens                   & 0.33                   & 0.00033                       & 1/300                                      & 11/10000000        & 0.99               \\
                           & S. Island                & 0.01                   & 0.00001                       & 0.1/100                                    & 1/100000000        & 0.01               \\
    \midrule
    \multirow{2}{*}{$r_6$} & Queens                   & 0.16                   & 0.00016                       & 1/300                                      & 1/1875000          & 0.93               \\
                           & S. Island                & 0.04                   & 0.00004                       & 0.1/100                                    & 1/25000000         & 0.07               \\
    \bottomrule
  \end{tabular}
  \caption{Candidate Generation State}
  \vspace{-15pt}
  \label{tbl:CandGen}
\end{table*}

\subsection{Phase 2: Candidate Probabilities Estimation}
\label{subsec:Probability}

This phase aims to estimate the probability of each candidate value to be the correct one. Current data cleaning systems do so by adopting various statistical methods. One typical method is the {\em NaiveBayes}~\cite{NaiveBayes}, where the probability that a cell $C$ of record $R$ has a certain candidate value $v$ ({\em Prob}~($C=v$)) is computed as {\em Prob}~($v \in D$) $\times$ $\prod_{A}$ {\em Prob}~(($v$ $\rightarrow$ $R.A$) $\in$ $D$), which is the probability of having $v$ in the whole dataset $D$, multiplied by, for each attribute $A$ other than $C$, the ratio of records in $D$ where having $v$ in $C$ implies the value in attribute $A$. Apparently this is not applicable to spatial data as the co-occurrence of $v$ and the spatial attributes in $A$ is pretty rare, which will make the probability of each candidate zero.

As it is the case for Phase~1, \sys enriches existing candidate probability estimators with spatial awareness. For any record $r_i$ with erroneous {\em Borough} cell, \sys modifies the candidate probability estimation in two ways: (1)~In the calculation of {\em Prob}~($C=v$), \sys replaces the term {\em Prob}~(($v$ $\rightarrow$ $R.A$) $\in$ $D$) by the term {\em Prob}~(($v$ $\rightarrow$ location near $R$) $\in$ $D$) for the spatial attributes in $A$. This means that instead of considering the exact co-occurrence between $v$ and each single location attribute, \sys employs its {\em spatial neighborhood} concept to consider the nearby co-occurrence between $v$ and $R$ according to the spatial proximity defined in the denial constraint. (2)~When calculating the co-occurrence probability, \sys does not count the nearby co-occurrences. Instead, it sums the co-occurrence weights as closer ones are weighted higher than further ones, per the {\em distance weighting} concept. Both nearby co-occurrences and their weights are directly obtained from the {\em DistanceMatrix}. Formally, in \sys, the probability estimation for any candidate value $v$ of cell $C$ of record $R$ is:

\vspace{-5pt}
\begin{equation}
    \begin{aligned}
        \label{eq:prob}
        Prob(C=v) = Prob(v \in D) \times \prod_{A'} Prob((v \rightarrow R.A') \in D) \\
        \qquad \times  Prob((v \rightarrow location \enspace near \quad R) \in D)
    \end{aligned}
\end{equation}
\vspace{-5pt}

, where $A'$ is the set of attributes in $R$ excluding $C$ and the spatial attributes. The first probability factor, {\em Prob}~($v \in D$), is basically $\frac{Count(v,D)}{|D|}$, which is the number of times that $v$ has appeared in the dataset $D$ divided by the number of records in $D$. The second probability factor for each attribute in $A'$, {\em Prob}~(($v$ $\rightarrow$ $R.A'$) $\in$ $D$), is $\frac{Count((v,R.A'),D)}{Count(v,D)}$, which is the number of times that $v$ and the value of $R.A'$ have appeared together in $D$ divided by the number of times that $v$ has appeared in $D$. The third probability factor, {\em Prob}~(($v$ $\rightarrow$ location near $R$) $\in$ $D$) is $\frac{|Spatial(v,R)|}{Count(v,D)}$, which is the sum of the weights of those records where $v$ appears within a spatial proximity of $R$, divided by the number of times that $v$ has appeared in $D$. With this, Equation~\ref{eq:prob} can be rewritten as:

\vspace{-5pt}
\begin{align*}
    \label{eq:probfrac}
    Prob(C=v) & = \frac{Count(v,D)}{|D|} \times \prod_{A'} \frac{Count((v,R.A'),D)}{Count(v,D)} \times  \frac{|Spatial(v,R)|}{Count(v,D)} \\
              & =\frac{|Spatial(v,R)|}{|D|} \times \prod_{A'} \frac{Count((v,R.A'),D)}{Count(v,D)}
\end{align*}
\vspace{-5pt}

, where $|${\em Spatial(v,R)}$|$ is basically the third column of Table~\ref{tbl:CandGen}, computed at Phase~1 (Section~\ref{subsec:Generation}). {\em Count}($v$,$D$) is obtained directly from the value frequency table computed in Figure~\ref{fig:DMExample}b. For {\em Count}(($v$,$R.A'$),$D$), we query the original input table to get the {\em count} of the number of $R.A'$ when the value of cell $R.C$ is $v$. In case $R.A'$ is the record identifier, such {\em count} would be set as either (a)~1, if the candidate value $v$ is the original value of $R.C$, as the record identifier would naturally appear only once in the dataset, combined with the record original value, regardless of whether it is right or not, or (b)~0.1, if $v$ is not the original value of $R.C$. Even though there is a zero co-occurrence between the record identifier and any value of $v$ that is not the original, we follow the same practice used by state-of-the-art data cleaning systems~\cite{GGM+22,RCI+17,WZI+20}, known as the principle of minimality, where we put 0.1 co-occurrence value for any non co-occurrence. This gives ten times more bias towards the original record value, which again, follows existing data cleaning systems that favor the original value.

\vspace{1pt}
\noindent\textbf{Example.} The fourth and fifth columns of Table~\ref{tbl:CandGen} give the two terms used to compute the probability of each candidate value for the records $r_1$ to $r_6$ from the example of Figure~\ref{fig:DMExample}. The sixth column in the table presents the probability of each candidate to be the correct value for the record, which is basically the multiplication of the fourth and fifth columns. For $r_1$, as a representative example, the first probability term (fourth column) is basically the sum of weights (third column) divided by 1,000, which is the total number of input records. Since our toy example has only one non-spatial attribute, namely, the record identifier, the probability of the value Manhattan and Queens for $r_1$ would need to be multiplied by 0.1/300. The 0.1 is the default minimal value and 300 is the number of times that the values Manhattan and Queens appear in the input dataset. Meanwhile, as Staten Island is the original value of $r_1$, we multiply the probability by 1/100, where 1 for the original value and 100 is the number of times of Staten Island in the input dataset. The output of this phase is both the second and sixth columns of Table~\ref{tbl:CandGen} as the candidate values with their probabilities.

\subsection{Phase 3: Candidate Labeling and Cutoffs}
\label{subsec:Labeling}

The goal of this phase is to identify: (a)~If there are candidate values that have marginal probability to the extent that there is no need to consider them further, and (b)~If there is a certain candidate value that is clearly dominant and we can safely identify that this is the correct value for its corresponding erroneous cell. To do so, as outlined in Algorithm~\ref{alg:CandGen}, \sys first normalizes the candidate probabilities to have a sum of~1. Then, it employs two parameters, {\em MinProb} and {\em MaxProb}. Any candidate value that has a probability less than {\em MinProb} will be considered marginal, removed, and not considered further. Then, if there is only one remaining candidate value or if there is one candidate value that has a significantly high probability more than {\em MaxProb}, we consider that this value is the correct one and move the corresponding cell from the erroneous list to the clean list. The output of this phase is the module output, which includes both the erroneous and clean cells, along with a list of remaining candidate values per each cell.

\vspace{2pt}
\noindent\textbf{Example.} The last column of Table~\ref{tbl:CandGen} gives the normalized probability of the sixth column. Assuming {\em MinProb}=0.05, we exclude the candidate value Queens from $r_2$, Manhattan from $r_4$, and Staten Island from $r_5$. Assuming {\em MaxProb}=0.95, we mark $r_5$.{\em Borough} as clean cell with value Queens. Clean and erroneous cells with their remaining candidate values are passed to the next module.

\section{Spatial Input Formulator}
\label{sec:InputFormulator}

The input to this module is the two sets of clean and erroneous cells, each with its own candidate list, identified from the {\em spatial candidate generation} module. Then, the module mainly operates on the erroneous cells, and aims to identify, with a score, the correct value for each erroneous cell. The score will be based on how much each candidate value satisfies (or violates) the spatial denial constraint. This means that if we have only one spatial constraint as in the example of Figure~\ref{fig:NYCExampleMap}, then the output of this module will be basically the output of the host data cleaning system, with a completely repaired dataset. However, \sys acknowledges that there could be other non-spatial constraints that would affect the final repaired value. Hence, the {\em Spatial Input Formulator} module will basically pass its findings to the {\em Error Correction} module of its host data cleaning systems. As \sys aims to be a generic framework that can inject spatial awareness in (any) data cleaning system, it needs to formulate its findings in the same format of its host system. In this section, we show how \sys can inject both the {\em Spatial Neighborhood} and {\em Distance Weighting} concepts in the input formulation module of three data cleaning systems, namely, AimNet~\cite{WZI+20} (Section~\ref{subsec:ViolationFeatureVector}), Baran~\cite{MA20} (Section~\ref{subsec:ProbabilityFeatureVector}), and HoloClean/MLNClean~\cite{GGM+22,RCI+17} (Section~\ref{subsec:FactorGraph}).

\begin{figure}[t]
    \centering
    \includegraphics[width=0.8\linewidth]{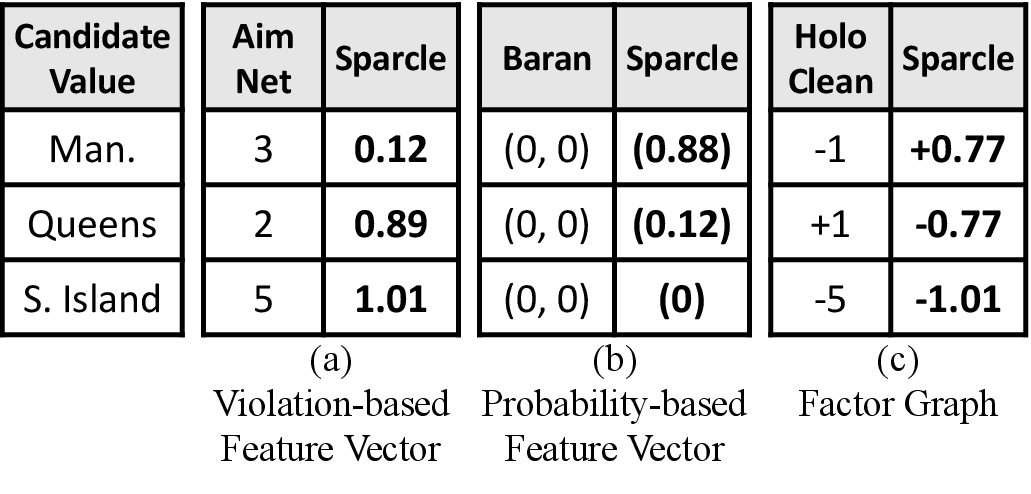}
    \vspace{-8pt}
    \caption{Input of $r_1.${\em \textmd{Borough}} by Original Method VS \sys}
    \vspace{-8pt}
    \label{fig:InputExample}
\end{figure}

\subsection{Violation-based Feature Vectors}
\label{subsec:ViolationFeatureVector}

AimNet~\cite{WZI+20}, the error correction method of the HoloClean's open source distribution, requires a feature vector $V$ per cell per constraint, where $v[i]$ represents the score of how the $i$th candidate of the cell violates the denial constraint. To construct its feature vector, AimNet~\cite{WZI+20} counts the number of violations of the constraint that are caused by the cell taking its $i$th candidate. The first column of Figure~\ref{fig:InputExample}a shows AimNet feature vector for $r_1$ in our running example of Figure~\ref{fig:DMExample}. Setting $r_1$.{\em Borough} to Manhattan will cause three instances of denial constraint violations with $r_4$, $r_5$, and $r_6$ (the 3rd to 5th rows in {\em DistanceMatrix}). Hence, the score is set to 3. Similarly, the scores for Queens and Staten Island are set to 2 and 5.

Apparently, this is not suitable for spatial constraints as it equally weights the constraint violations between $r_1$ and $r_2$ with the constraint violation between $r_1$ and $r_5$. Spatially speaking, the constraint violation ($r_1$, $r_5$) should be {\em weaker} than ($r_1$, $r_2$), as the distance between $r_1$ and $r_5$ is much more than the distance between $r_1$ and $r_2$. To inject the spatial awareness in the input formulator of AimNet, \sys fills in the feature value $V[i]$ by summing up the weights of violations that are caused by the cell taking its $i$th candidate. The second column of Figure~\ref{fig:InputExample}a gives such vector for $r_1$. If $r_1$.{\em Borough} is set to Manhattan, then three constraint violations would take place with $r_4$, $r_5$, and $r_6$, each with a weight 0.04, and hence the total score of the constraint violation would be 0.12. Similar, if $r_1$.{\em Borough} is set to Queens, then two constraint violations would take place with $r_2$ and $r_3$, with weights 0.64 and 0.25, respectively. Hence, the total score is 0.89. Finally, if $r_1$.{\em Borough} is set to Staten Island, then five constraint violations would take place with $r_2$ to $r_6$ with a total weight of 1.01.

It is important to note here that the lower the score the more likely the value would be considered correct. The fact that Manhattan has the lowest score of 0.12, i.e., the lowest violation score, is an indication that, per the spatial constraint, it is the most favored value for $r_1$.{\em Borough}. The final feature vector from \sys will be passed to the repair module of AimNet~\cite{WZI+20} to consider it along with other non-spatial constraints for the final repaired dataset. 

\subsection{Probability-based Feature Vectors}
\label{subsec:ProbabilityFeatureVector}

Unlike AimNet~\cite{WZI+20}, the Baran~\cite{MA20} system requires the input to its error correction method as a feature vector per cell per candidate, where each vector value represents the probability of the candidate according to a specific dependency. Meanwhile, Baran does not ask for user-input constraints, instead, it assumes all possible dependencies from all other attributes to the cell. For the example of Figure~\ref{fig:DMExample}, Baran assumes the dependency from {\em Latitude} and {\em Longitude} separately. Then, for each dependency, e.g., {\em Latitude} $\rightarrow$ {\em Borough}, it estimates the probability based on co-occurrence. The first column in Figure~\ref{fig:InputExample}b gives the feature vector values for $r_1$ for each possible {\em Borough} value. They are all zero vectors as there is zero co-occurrence between the {\em Latitude} and {\em Longitude} values with any {\em Borough} value. This makes the error correction module of Baran fail to identify the correct answer.

To inject the spatial awareness in the input formulator of Baran, \sys uses the weights and spatial neighborhood records that were evaluated in Section~\ref{sec:ErrorDetector} along with the candidate values computed in Section~\ref{sec:CandidateGenerator} to calculate the probability of a combined dependency on the form {\em (Latitude, Longitude)} $\rightarrow$ {\em Borough}. The second column of Figure~\ref{fig:InputExample}b gives such vector for $r_1$. Since there is only two possible values among the candidate ones that have proximity co-occurrence with $r_1$, we set their vector values as their normalized probability 0.88 for Manhattan and 0.12 for Queens. The last row for Staten Island is set to 0 as there is no proximity co-occurrence.

Unlike the case of AimNet, the higher the values here the more likely the candidate value is the correct one. This is mainly because these values represent a probability rather than a violation. Finally, such form of output vector of \sys will be sent to the repair module of Baran~\cite{MA20} to consider it along with other non-spatial constraints for the final repaired dataset.

\subsection{Factor Graph}
\label{subsec:FactorGraph}

HoloClean~\cite{RCI+17} and MLNClean~\cite{GGM+22} include error correction methods that are based on Markov Logic Network~\cite{RD06}, which requires its input to be in a form of a factor graph. To construct the factor graph, each functional dependency instance needs a factor function that returns a value reflecting how the instance satisfies the dependency. In particular, for HoloClean, the factor function returns 1 if the instance satisfies the dependency, otherwise it returns $-1$. Then, the data cleaning process aims to find the values that maximize the sum of all factor functions in the dataset. The first column in Figure~\ref{fig:InputExample}c shows the sum of factor functions related with $r_1$ in Figure~\ref{fig:DMExample}. 
For Manhattan, the sum of factor functions would be -1 as the factor function would return -1 three times (with $r_4$, $r_5$ and $r_6$) and 1 two times (with $r_2$ and $r_3$) when $r_1$.{\em Borough} is set to Manhattan.
For Staten Island, the sum would be -5 as the factor function would return -1 five times.

Apparently, this is not suitable for spatial constraints as the factor function would only return either 1 or -1 regardless of how strong a certain instance satisfies the spatial constraint. To inject the spatial awareness in the input formulator of HoloClean~\cite{RCI+17}, \sys modifies the factor graph construction by multiplying the factor function output (1 or -1) by the weight of the instance, computed in the {\em DistanceMatrix} of  Figure~\ref{fig:DMExample}. For example, the second column in Figure~\ref{fig:InputExample}c gives \sys output for factor graphs for $r_1$. 
For Manhattan, the sum of factor functions would be 0.77, computed as -1*(0.04+0.04+0.04)+1*(0.64+0.85). 
For Queens, the sum of factor functions would be 1*(0.04+0.04+0.04)-1*(0.64+0.85)=-0.77. 
For Staten Island, the sum will be -1.01. 
The higher the value of the sum the more likely the candidate value is the correct one. This shows how spatial awareness changed the favored value from Queens to Manhattan. Such form of factor graph of \sys will be sent to the repair module of HoloClean~\cite{RCI+17} to consider it along with other non-spatial constraints for the final repaired dataset.

\section{Experimental Results}
\label{sec:Experiment}

\begin{table}[t]
   \small
    \setlength{\tabcolsep}{3pt}
    \centering
    \begin{tabular}{llrrrr}
        \toprule
        \textbf{Dataset}                                                             & \textbf{Dependency}                              & \textbf{Records}        & \textbf{Errors} & \textbf{Dup.} & \textbf{Dis.} \\
        \midrule
        \multirow{2}{*}{\begin{tabular}[c]{@{}l@{}}Austin-\\Code\end{tabular}}       & ({\em Lat}, {\em Lon}) $\rightarrow$ {\em zipcode         } & \multirow{2}{*}{93,414}    & 13,968             & 0.00                & 50                          \\
                                                                                     & ({\em Lat}, {\em Lon}) $\rightarrow$ {\em city            } &                            & 12,224             & 0.00                & 9                           \\
        \midrule
        \multirow{3}{*}{\begin{tabular}[c]{@{}l@{}}Chicago-\\Building\end{tabular}}  & ({\em Lat}, {\em Lon}) $\rightarrow$ {\em community } & \multirow{3}{*}{731,734}   & 105,240            & 0.64                & 77                          \\
                                                                                     & ({\em Lat}, {\em Lon}) $\rightarrow$ {\em census   } &                            & 138,953            & 0.64                & 980                         \\
                                                                                     & ({\em Lat}, {\em Lon}) $\rightarrow$ {\em ward            } &                            & 181,119            & 0.58                & 50                          \\
        \midrule
        \multirow{2}{*}{\begin{tabular}[c]{@{}l@{}}NYC-\\Crash\end{tabular}}         & ({\em Lat}, {\em Lon}) $\rightarrow$ {\em borough         } & \multirow{2}{*}{1,751,624} & 421,013            & 0.44                & 5                           \\
                                                                                     & ({\em Lat}, {\em Lon}) $\rightarrow$ {\em zipcode         } &                            & 528,565            & 0.30                & 230                         \\
        \midrule
        \multirow{5}{*}{\begin{tabular}[c]{@{}l@{}}Chicago-\\Synthetic\end{tabular}} & ({\em Lat}, {\em Lon}) $\rightarrow$ {\em district} &                            &                    &                     & 23                          \\
                                                                                     & ({\em Lat}, {\em Lon}) $\rightarrow$ {\em ward}             &                            &                    &                     & 50                          \\
                                                                                     & ({\em Lat}, {\em Lon}) $\rightarrow$ {\em zipcode        }  &                            &                    &                     & 59                          \\
                                                                                     & ({\em Lat}, {\em Lon}) $\rightarrow$ {\em beat            } &                            &                    &                     & 275                         \\
                                                                                     & ({\em Lat}, {\em Lon}) $\rightarrow$ {\em census   } &                            &                    &                     & 801                         \\
        \bottomrule
    \end{tabular}
    \caption{Experiment Datasets}
    \vspace{-15pt}
    \label{tbl:ExperimentData}
\end{table}

\begin{figure*}[t]
\centering
\subfigure[Range threshold d (m)]{
\label{fig:exp:AccR}
\includegraphics[width=0.23\linewidth, height=2.5cm]{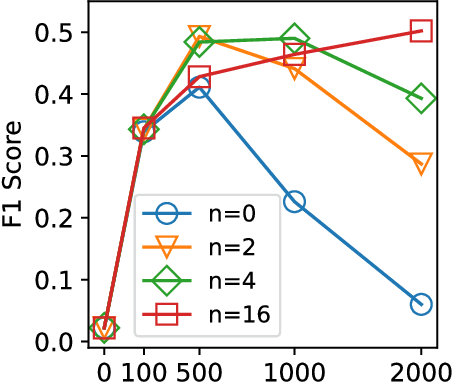}
}
\subfigure[Range threshold d (m)]{
\label{fig:exp:EffR}
\includegraphics[width=0.23\linewidth, height=2.5cm]{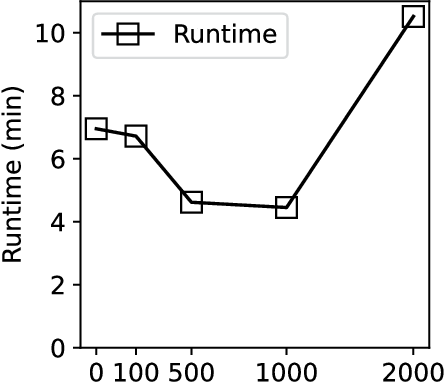}
}
\subfigure[kNN threshold k]{
\label{fig:exp:AccK}
\includegraphics[width=0.23\linewidth, height=2.5cm]{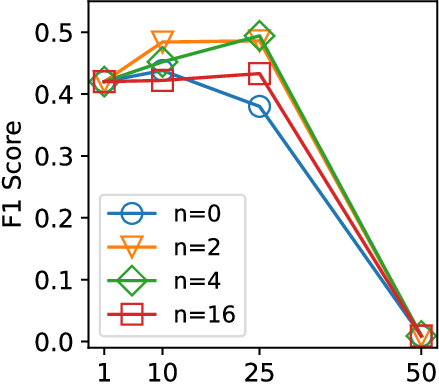}
}
\subfigure[kNN threshold k]{
\label{fig:exp:EffK}
\includegraphics[width=0.23\linewidth, height=2.5cm]{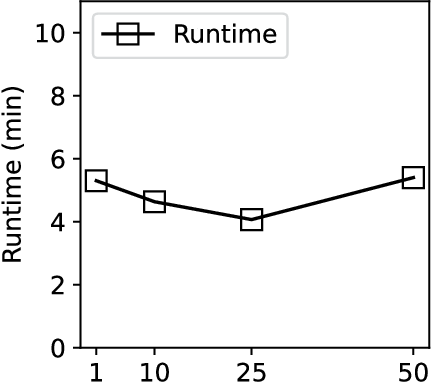}
}
\vspace{-5pt}
\caption{\sys Parameter Tuning}
\vspace{-5pt}
\label{fig:exp:tuning}
\end{figure*}

This section compares the performance of \sys based on a real system implementation inside the open-source distribution of the HoloClean system~\cite{HoloClean}, which has the AimNet~\cite{WZI+20} as its error correction method. All experiments are performed on a Linux server with 20 CPU@2.2GHz, 96GB memory and 1TB SSD. The implementation and experiment code are open source~\cite{Sparcle}. For the {\em spatial neighborhood} concept, we use both range and $k$NN spatial constraints as defined in Section~\ref{subsec:Constraints}. For the {\em distance weighting} concept, we use the weight function $\mathcal{W}(r_i, r_j)$ = $(1-\frac{\mathcal{F}(r_i, r_j)}{d})^n$, where $\mathcal{F}(r_i, r_j)$ is the distance between records $r_i$ and $r_j$, and $n$ is an exponential weight parameter, where larger $n$ gives more weight to closer records. For spatial range constraints, $d$ is the range threshold defined in the denial constraints. For $k$NN constraints, $d$ is the distance of the $k$th nearest neighbor from $r_i$.

We compare \sys against two recent state-of-the-art rule-based data cleaning systems (1)~HoloClean~\cite{RCI+17}, based on its open-source distribution~\cite{HoloClean}, which is a data repairing system that unifies integrity constraints with others statistical signals. We mute all signals other than integrity constraints to limit the comparison to constraint-based data cleaning. (2)~Baran~\cite{MA20}, which is a configuration-free error correction system that assumes dependencies between every pair of attributes, while using the Raha system~\cite{MAF+19} for its error detection. In our comparison, we show the performance for two versions of \sys: (a)~our default \sys implementation with an exponential weight parameter $n$=2, and (b)~\sys with $n$=0, which basically cancels the {\em distance weighting} concept as the weight between any pair of records would be always 1 regardless of how far they are from each other. The main purpose of having such version of \sys is to have kind of an ablation study that shows the impact of the {\em distance weighting} concept on the overall performance of \sys.

\vspace{2pt}
\noindent\textbf{Datasets.} Table~\ref{tbl:ExperimentData} shows the properties of the datasets we are using in our experiments. We use the following three real datasets and one synthetic data, where for each real dataset, we only keep the spatial attributes {\em Latitude} and {\em Longitude} and the attributes dependent on them: (1)~Austin-Code~\cite{AustinOpenData}. This is 93+K records for the locations of Austin Code Department complaint cases over the last 8 years. Two spatial functional dependencies need to be kept for this data, ({\em Lat}, {\em Lon}) $\rightarrow$ {\em zipcode} and ({\em Lat}, {\em Lon}) $\rightarrow$ {\em city}. There are around 14K and 12K records violating the first and second constraints, respectively. The second last column in Table~\ref{tbl:ExperimentData} presents the error duplication ratio, which is the ratio of erroneous records that took place on the same location of some other records. The last column represents the number of distinct values for the dependent attributes (i.e., zipcode and city). (2)~Chicago-Building~\cite{ChicagoBuilding}. This is 731+K records for the currently-valid building permits issued by the City of Chicago since 2006. The dataset has three spatial functional dependencies, where around 105K, 139K, and 181K records are erroneous for each of the functional dependencies. (3)~NYC-Crash~\cite{NYCOpenData}. This is the dataset shown in Figure~\ref{fig:NYCExample} with 1.7+M records for the location of vehicle crashes in NYC since 2014, with two functional dependencies. 421K and 528K are violating the first and second dependency. (4)~Chicago-Synthetic. This is the synthetic data in the spatial extent of Chicago, Illinois, in which each record is a randomly generated location. We include five functional dependencies to this dataset. We are not including the number of records, errors, and duplication ratio for this dataset, as these are parameters that we would vary in our experiments. We determine the ground truth for each dataset by referring to the corresponding government-issued boundaries.

\vspace{2pt}
\noindent\textbf{Evaluation Metrics and Experiment Design}. We evaluate \sys and its competitors based on four metrics: (1)~{\em Precision}: The fraction of number of correct repairs over total number of repairs made by the system, (2)~{\em Recall}: The fraction of number of correct repairs over total number of errors, (3)~{\em F1 score}: The harmonic mean of precision and recall, i.e., $\frac{2*(Prec. * Rec.)}{Prec. + Rec.}$, and (4)~the system run time. In this section, we first perform a parameter study of \sys to set on its optimal parameters (Section~\ref{subsec:ParameterStudy}). Then, we compare \sys against competitors in terms of accuracy  (Sections~\ref{subsec:Accuracy} to~\ref{subsec:Distinct}) and efficiency (Section~\ref{subsec:Efficiency}).


\subsection{\sys Parameter Tuning}
\label{subsec:ParameterStudy}

This section studies the impact and trade-offs of \sys parameters on its accuracy and efficiency, namely, the spatial range $d$, the number of nearest neighbors $k$, and the exponential weight parameter $n$. To do so, we create a Chicago-Synthetic dataset of 20,000 records with schema ({\em Lat}, {\em Lon}, {\em census}) and focus on the functional dependency ({\em Lat}, {\em Lon}) $\rightarrow$ {\em census}. The data has no duplicate locations and has 2,000 errors in the {\em census} attribute.

\vspace{1pt}
\noindent\textbf{Spatial range parameter $d$}. Figures~\ref{fig:exp:AccR} and~\ref{fig:exp:EffR} show the impact of increasing the spatial range $d$ from 0 to 2000, on both the accuracy ({\em F1 score}) and efficiency, respectively. For accuracy (Figure~\ref{fig:exp:AccR}), we plot \sys with different values of $n$ as it impacts the system accuracy. No need to do the same for efficiency as $n$ has no impact on the system efficiency. Note that setting $d$ to 0 is equivalent to not considering spatial awareness at all, in which \sys will perform as poorly as its host data cleaning system HoloClean. 
For $d$ from 100 to 2000, the highest {\em F1 score} at each distance increase, and finally, \sys achieves its highest {\em F1 score} when $d$=2000 with $n$=16. This suggests that a larger neighborhood has a potential to achieve higher accuracy. 
Meanwhile for $n$, a larger neighborhood requires a larger $n$ to achieve that high accuracy. For example for $d$ of 500, 1000 and 2000, the best $n$ is 2, 4, 16, respectively. Note that $n$=0 cancels the {\em distance weighting} concept, and hence gives the lowest accuracy. However, from efficiency perspective (Figure~\ref{fig:exp:EffR}), large value of $d$ (e.g., 2000) encounters very high overhead. In fact, there are two factors controlling the system efficiency, the learning time in the error correction method (i.e., AimNet) and the {\em DistanceMatrix} computation time in Section~\ref{subsec:DataStructure}. For smaller $d$, there is no much records within the spatial neighborhood, hence, a large portion of the data is deemed clean. With too much clean data, the learning time becomes the dominating factor in efficiency. So, larger $d$ would mean less learning time and hence better efficiency. However, at some point, larger $d$ would make the {\em DistanceMatrix} computation time the dominating factor, and then increasing $d$ would result in lower efficiency. Overall, in this example having $d=1000$ with $n=4$ achieves the best trade-off.

\vspace{1pt}
\noindent\textbf{Nearest-neighbor parameter $k$}. Figures~\ref{fig:exp:AccK} and~\ref{fig:exp:EffK} show the impact of increasing the $k$ from 1 to 50, on both the accuracy ({\em F1 score}) and efficiency, respectively. Similar to the analysis for the case of spatial range parameter $d$, a larger neighborhood in company with a larger value of $n$ achieves high accuracy. However, the neighborhood cannot be too large. In particular, the accuracy almost drops to 0 when $k$=50. The reason is that the dataset has 20,000 records but 801 distinct values of the {\em census}. So, a pretty large neighborhood (e.g., $k$=50) would involve too many neighbors from different census\_tracts, which may not be able to guide the cleaning logic. Hence, we come up with a recommendation ceiling value for $k$ that $k$ should not be greater than $\frac{|D|}{|A|}$ where $|D|$ is the size of the dataset and $|A|$ is the number of distinct values of attribute $A$. As for efficiency (Figure~\ref{fig:exp:EffK}), it is also similar to the case of spatial range that smaller value of $k$ will suffer from a dominating learning time, and higher values of $k$ suffer from the {\em DistanceMatrix} computation.

\begin{table*}[t]
    \setlength{\tabcolsep}{2pt}
    \begin{tabular}{lr|cccc|cccc|cccc}
        \toprule
        \multicolumn{2}{c|}{}                                                       & \multicolumn{4}{c|}{Precision} & \multicolumn{4}{c|}{Recall}                           & \multicolumn{4}{c}{F1 Score}                                                                                                                                                                                                                                                                                                                                     \\
        \midrule
        Dataset                                                                     & Attribute                      & {\begin{tabular}[c]{@{}c@{}}\sys\\(n=2)\end{tabular}} & {\begin{tabular}[c]{@{}c@{}}\sys\\(n=0)\end{tabular}} & HoloClean & Baran          & {\begin{tabular}[c]{@{}c@{}}\sys\\(n=2)\end{tabular}} & {\begin{tabular}[c]{@{}c@{}}\sys\\(n=0)\end{tabular}} & HoloClean & Baran    & {\begin{tabular}[c]{@{}c@{}}\sys\\(n=2)\end{tabular}} & {\begin{tabular}[c]{@{}c@{}}\sys\\(n=0)\end{tabular}} & HoloClean & Baran    \\
        \midrule
        \multirow{2}{*}{\begin{tabular}[c]{@{}l@{}}Austin-\\Code\end{tabular}}      & zipcode                        & \textbf{0.853}                                        & 0.790                                                 & 0.001     & 0.374          & \textbf{0.782}                                        & \textbf{0.782}                                        & 0.000     & 0.010    & \textbf{0.816}                                        & 0.786                                                 & 0.000     & 0.019    \\
                                                                                    & city                           & 0.992                                                 & 0.992                                                 & 0.993     & \textbf{0.995} & \textbf{0.992}                                        & \textbf{0.992}                                        & 0.024     & 0.674    & \textbf{0.992}                                        & \textbf{0.992}                                        & 0.046     & 0.683    \\
                                                                                    & Overall                        & \textbf{0.921}                                        & 0.885                                                 & 0.093     & 0.882          & \textbf{0.880}                                        & \textbf{0.880}                                        & 0.011     & 0.324    & \textbf{0.900}                                        & 0.882                                                 & 0.020     & 0.441    \\
        \midrule
        \multirow{3}{*}{\begin{tabular}[c]{@{}l@{}}Chicago-\\Building\end{tabular}} & community                      & \textbf{0.990}                                        & 0.983                                                 & 0.925     & -$^{*}$        & \textbf{0.982}                                        & 0.977                                                 & 0.635     & -$^{*}$  & \textbf{0.986}                                        & 0.980                                                 & 0.753     & -$^{*}$  \\
                                                                                    & census\_tract                  & \textbf{0.829}                                        & 0.724                                                 & 0.398     & -$^{*}$        & \textbf{0.894}                                        & 0.893                                                 & 0.393     & -$^{*}$  & \textbf{0.860}                                        & 0.800                                                 & 0.396     & -$^{*}$  \\
                                                                                    & ward                           & \textbf{0.746}                                        & 0.727                                                 & 0.717     & -$^{*}$        & 0.685                                                 & \textbf{0.710}                                        & 0.437     & -$^{*}$  & 0.714                                                 & \textbf{0.718}                                        & 0.543     & -$^{*}$  \\
                                                                                    & Overall                        & \textbf{0.836}                                        & 0.785                                                 & 0.627     & -$^{*}$        & 0.827                                                 & \textbf{0.836}                                        & 0.472     & -$^{*}$  & \textbf{0.832}                                        & 0.810                                                 & 0.538     & -$^{*}$  \\
        \midrule
        \multirow{2}{*}{\begin{tabular}[c]{@{}l@{}}NYC-\\Crash\end{tabular}}        & borough                        & \textbf{0.998}                                        & 0.997                                                 & 0.683     & -$^{\#}$       & \textbf{0.994}                                        & \textbf{0.994}                                        & 0.587     & -$^{\#}$ & \textbf{0.996}                                        & 0.995                                                 & 0.632     & -$^{\#}$ \\
                                                                                    & zipcode                        & \textbf{0.821}                                        & 0.803                                                 & 0.384     & -$^{\#}$       & 0.662                                                 & \textbf{0.667}                                        & 0.264     & -$^{\#}$ & \textbf{0.733}                                        & 0.729                                                 & 0.313     & -$^{\#}$ \\
                                                                                    & Overall                        & \textbf{0.909}                                        & 0.898                                                 & 0.533     & -$^{\#}$       & 0.809                                                 & \textbf{0.812}                                        & 0.407     & -$^{\#}$ & \textbf{0.856}                                        & 0.853                                                 & 0.462     & -$^{*}$  \\
        \bottomrule
    \end{tabular}
    \begin{flushleft}
        {\small * cannot finish due to memory error in error correction}

        \vspace{-3pt}

        {\small $\#$ cannot finish after 1 day}
    \end{flushleft}
    \caption{Cleaning Accuracy on Real Data}
    \vspace{-10pt}
    \label{tbl:Accuracy}
\end{table*}

\subsection{System Overall Accuracy}
\label{subsec:Accuracy}

Table~\ref{tbl:Accuracy} gives the cleaning accuracy for two versions of \sys (with $n$=2 and $n$=0), compared to HoloClean and Baran systems. The accuracy is listed in terms of precision, recall, and F1 score for each functional dependency, and then for the overall accuracy for the whole dataset with all functional dependencies combined. Notice that the overall accuracy is not the average accuracy over all dependencies, as some records have more than one functional dependency corrected, and then it was counted in each of them. The overall accuracy is computed based on records that are completely corrected for all their functional dependencies. For Baran system, we only show the results for the Austin dataset as Baran did not scale up to run on larger datasets. More about this is in Section~\ref{subsec:Efficiency}. For the F1 score, \sys clearly outperforms HoloClean and Baran for every single dependency accuracy as well as the overall accuracy. The most notable results are for Austin dataset, where both HoloClean and Baran perform extremely poor. In particular, for the {\em zipcode} dependency, HoloClean and Baran have F1 scores of 0 and 0.019, respectively, compared to 0.816 for \sys. The main reason here is that, as depicted in Table~\ref{tbl:ExperimentData}, Austin data has no duplicates, which makes it very hard to clean by current data cleaning systems. Meanwhile, even though the {\em city} dependency also has no duplicates, but HoloClean and Baran have better performance for it than the {\em zipcode} dependency with F1 scores of 0.046 and 0.683, though still way worse than \sys with F1 score of 0.992. The main reason is that the {\em city} dependency has only 9 distinct values, which is much less than the distinct values for {\em zipcode}, which is 50 (Table~\ref{tbl:ExperimentData}). Naturally, it is much harder to clean data with more distinct values. For \sys, the impact of $n$=2 over $n$=0 appears more with higher duplicate ratios and large number of distinct values. Generally speaking, there are two main factors that control the accuracy of data cleaning systems, the duplication ratio and number of distinct values. Hence, Sections~\ref{subsec:DupRatio} and~\ref{subsec:Distinct} will discuss the impact of these two factors in more details.

The relative performance of all systems for the Recall is pretty similar to that of F1 score, where \sys clearly outperforms HoloClean and Baran for all dependencies and overall accuracy. Notice that the Recall results for the {\em Borough} dependency of NYC data is the one that was mentioned earlier in Table~\ref{tbl:NYCResult} for Figure~\ref{fig:NYCExample}. For the precision, both HoloClean and Baran tend to get higher values with small number of distinct values. The main reason is that it is somehow easier to guess the right value when trying to make a repair. For example, in the {\em city} dependency of the Austin data, most of the correct values for the erroneous entries is "Austin". Since Baran and HoloClean just tried to guess "Austin" as a correct value, they end up with a high precision value, even though their recall is pretty low, and hence they also have pretty low F1 score.

\begin{figure*}[t]
    \centering
    \includegraphics[width=\linewidth]{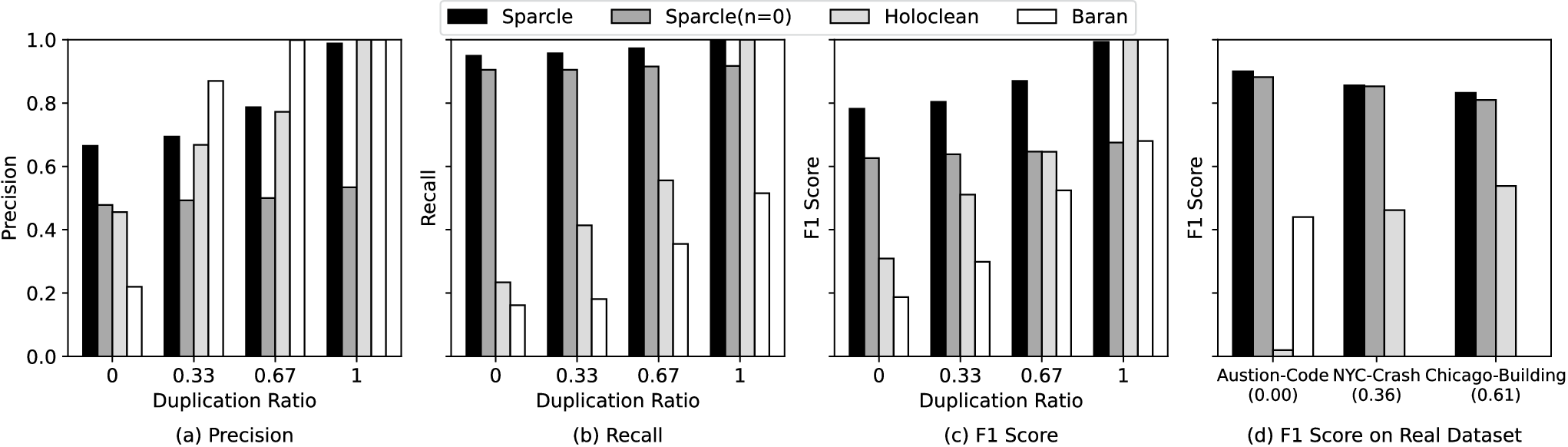}
    \vspace{-15pt}
    \caption{Impact of Duplication Ratio}
    \vspace{-10pt}
    \label{fig:DupRatio}
\end{figure*}

\subsection{Impact of Duplicate Ratio on Accuracy}
\label{subsec:DupRatio}

Figure~\ref{fig:DupRatio} shows the impact of duplication ratio on the accuracy of \sys, HoloClean, and Baran. To control the duplication ratio, we use the Chicago-Synthetic dataset (20K records with 2K errors) in Figure~\ref{fig:DupRatio}a to~\ref{fig:DupRatio}c, where we measure the precision, recall, and F1 score when having error duplication ratios of 0, 0.33, 0.67, and 1 for the functional dependency ({\em Lat}, {\em Lon}) $\rightarrow$ {\em ward}. A duplication ratio of 0 means that none of the erroneous records happen in a location of other records, while a duplication ratio of 1 means that all erroneous records happened in the same exact location of some other records. \sys with $n$=2 significantly outperforms HoloClean and Baran in low duplicate ratios for precision, recall, and F1 score, and gives similar performance for duplicate ratio of 1. The main reason is that having erroneous records with duplicate values gives HoloClean and Baran the chance to learn the correct values form the duplicates, and hence can perform better. However, \sys can still perform well even with 0 duplicate ratio, as it employs the spatial neighborhood concept, which somehow considers records with spatial proximity as duplicates. Meanwhile, \sys with $n$=2 performs much better than the case when $n$=0, and the superiority increases with high duplicate ratios. The main reason is that with more duplicates, it becomes more important to set accurate weights for each record with respect to satisfying the functional dependency. Hence, \sys with $n$=2 takes advantage of its distance weighting concept to accommodate this.

Figure~\ref{fig:DupRatio}d shows the same experiment of Figure~\ref{fig:DupRatio}c, yet for the three real datasets of Austin, NYC, and Chicago that have duplicate ratios 0, 0.36, and 0.61, respectively. For each dataset, we compute the duplicate ratio as a weighted average of the duplicate ratios of its dependencies. We could not run Baran for NYC and Chicago. For all cases, \sys significantly outperforms its competitors. HoloClean is doing extremely poor in Austin data that has 0 duplicate ratio.

\begin{table}[t]
    \setlength{\tabcolsep}{2pt}
    \centering
    \begin{tabular}{ll|cccc}
        \toprule
        \textbf{Attribute}                & \textbf{Metric} & {\begin{tabular}[c]{@{}c@{}}\sys\\(n=2)\end{tabular}} & {\begin{tabular}[c]{@{}c@{}}\sys\\(n=0)\end{tabular}} & HoloClean & Baran \\
        \midrule
        \multirow{3}{*}{police\_district} & Prec.           & \textbf{0.97}                                         & 0.95                                                  & 0.61      & 0.46  \\
                                          & Rec.            & \textbf{0.99}                                         & \textbf{0.99}                                         & 0.31      & 0.13  \\
                                          & F1              & \textbf{0.98}                                         & 0.97                                                  & 0.41      & 0.18  \\
        \midrule
        \multirow{3}{*}{ward}             & Prec.           & \textbf{0.67}                                         & 0.47                                                  & 0.46      & 0.40  \\
                                          & Rec.            & \textbf{0.95}                                         & 0.90                                                  & 0.23      & 0.06  \\
                                          & F1              & \textbf{0.78}                                         & 0.62                                                  & 0.31      & 0.08  \\
        \midrule
        \multirow{3}{*}{zipcode}          & Prec.           & \textbf{0.86}                                         & 0.79                                                  & 0.21      & 0.44  \\
                                          & Rec.            & \textbf{0.98}                                         & 0.97                                                  & 0.11      & 0.07  \\
                                          & F1              & \textbf{0.92}                                         & 0.87                                                  & 0.14      & 0.09  \\
        \midrule
        \multirow{3}{*}{beat}             & Prec.           & \textbf{0.60}                                         & 0.49                                                  & 0.43      & 0.60  \\
                                          & Rec.            & \textbf{0.93}                                         & 0.89                                                  & 0.21      & 0.01  \\
                                          & F1              & \textbf{0.73}                                         & 0.63                                                  & 0.28      & 0.2   \\
        \midrule
        \multirow{3}{*}{census\_tract}    & Prec.           & \textbf{0.35}                                         & 0.25                                                  & 0.03      & 0.23  \\
                                          & Rec.            & \textbf{0.84}                                         & 0.77                                                  & 0.01      & 0.00  \\
                                          & F1              & \textbf{0.49}                                         & 0.38                                                  & 0.02      & 0.01  \\
        \bottomrule
    \end{tabular}
    \caption{Accuracy per Attribute in Chicago-Synthetic.}
    \label{tbl:Synthetic}
    \vspace{-25pt}
\end{table}

\subsection{Impact of Distinct Values on Accuracy}
\label{subsec:Distinct}

\begin{figure*}[t]
    \centering
    \includegraphics[width=\linewidth]{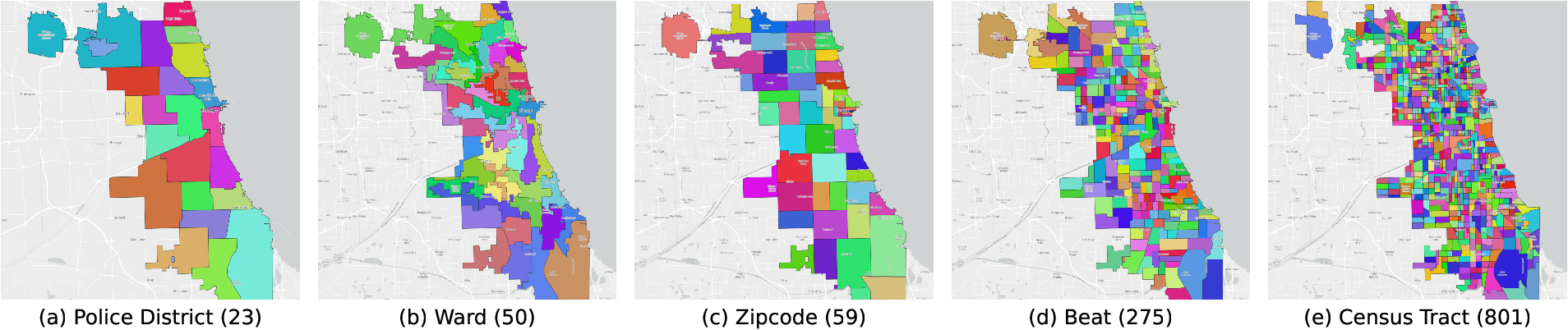}
    \vspace{-6pt}
    \caption{Map of Dependency Attribute of Chicago-Synthetic}
    \vspace{-6pt}
    \label{fig:FiveAttrMap}
\end{figure*}

To better study the impact of the number of distinct values, we use the Chicago-Synthetic dataset (20K records, 2K errors) with five different dependencies, each with different number of distinct values as outlined in Table~\ref{tbl:ExperimentData}. To visualize the challenges, Figure~\ref{fig:FiveAttrMap} plots the map of Chicago outlines the possible values of the five dependency attributes, {\em police\_district}, {\em ward}, {\em zipcode}, {\em city}, and {\em census\_tract}, with 23, 50, 59, 275, and 801 distinct values, respectively. Table~\ref{tbl:Synthetic} gives the precision, recall, and F1 score for all the five dependencies for \sys ($n$=2 and $n$=0), HoloClean and Baran. \sys clearly outperforms HoloClean and Baran in all measures and dependencies. A general trend is that the accuracy of all techniques degrade with the increase of the number of distinct values, yet, \sys is way more resilient to number distinct value than its competitors. This is to the extent that, for the census tract, the F1 score for HoloClean and Baran is 0.02 and 0.01, respectively, while it is 0.49 for \sys ($n$=2). This shows that \sys is still able to clean a good ratio of the records, while other system cannot really clean any record here. The main reason is that, as plotted in Figure~\ref{fig:FiveAttrMap}e for census tract boundaries, with more area distinct values, there are longer boundaries between each two values and more records close to the boundary, which are naturally hard to clean. 

It is interesting to see that though the {\em zipcode} dependency has slightly more distinct values than the {\em ward} dependency, but \sys actually performs better for it. The main reason is that per the boundary maps of Figures~\ref{fig:FiveAttrMap}b and~\ref{fig:FiveAttrMap}c, the {\em ward} areas have more complex shapes than {\em zipcode} areas, which means that there are longer boundaries between areas, and hence it is much harder to clean. Overall, the shape of the boundary areas as well as the number of areas (i.e., distinct value) impact the accuracy of all techniques. Finally, Table~\ref{tbl:Synthetic} shows the impact of the {\em distance weighting} concept in \sys, where having $n$=2 makes \sys way more resilient to larger numbers of distinct values. With $n$=0, the accuracy of \sys degrades much higher with the increase of distinct values.

\begin{table}[t]
    \centering
    \begin{tabular}{l|rrr}
        \toprule
        \textbf{Dataset} & \textbf{\sys} & HoloClean & Baran    \\
        \midrule
        Austin-Code      & 22m10s        & 17m39s    & 1h03m50s \\
        Chicago-Building & 2h14m29s      & 1h55m28s  & -$^{*}$  \\
        NYC-Crash        & 3h58m10s      & 3h06m55s  & -$^{\#}$ \\
        \bottomrule
    \end{tabular}
    \begin{flushleft}
        {\small * cannot finish due to memory error}

        \vspace{-5pt}

        {\small $\#$ cannot finish after 1 day}
    \end{flushleft}
    \caption{Running Time on Real Data}
    \vspace{-25pt}
    \label{tbl:Efficiency}
\end{table}

\subsection{System Efficiency}
\label{subsec:Efficiency}

Table~\ref{tbl:Efficiency} gives the running time for \sys, HoloClean, and Baran for the three real datasets. We did not include \sys with $n$=0 as the value of $n$ does not affect the system efficiency. As was shown earlier in Table~\ref{tbl:Accuracy}, Baran could not finish for Chicago and NYC datasets, while it takes too much time for the Austin dataset. The reason is twofold. First, Baran is proposed as an in-memory framework, so, it cannot handle out-of-memory large datasets. Second, Baran assumes functional dependency between every pair of attributes, hence ends up in processing too many dependencies, which takes significant execution time overhead. For all datasets, \sys encounters 17\% to 29\% extra overhead than HoloClean. This is mainly due to the time taken to build the {\em DistanceMatrix} as a self-join of the input data. Given that spatial operations are pretty expensive compared to non-spatial operations, having less than 30\% overhead is highly acceptable. This is mainly due to the {\em DistanceMatrix} that is computed in the first component of \sys (Section~\ref{sec:ErrorDetector}) to avoid excessive spatial operations in later components.

\section{Related Work}
\label{sec:relatedWork}

Motivated by a real need, there has been huge efforts over the last two decades to build systems and algorithms for automated data cleaning~\cite{IC15,IC19,IN22}. By far, the most common of these approaches are the rule-based data cleaning systems (e.g.,~\cite{DEE+13,GGM+22,KIJ+15,GMP+20,MA20,RCI+17,ROA+21}), where it is deployed in open-source systems~\cite{Baran,HoloClean,Llunatic,openclean} and industry~\cite{Inductiv,Tamr,Trifacta}. Almost all rule-based approaches share the core idea and goal of trying to minimize or eliminate a set of constraint violations, where a constraint is presented as a dependency rule that needs to be followed by any given dataset. Recent notable examples of such approaches include the HoloClean~\cite{RCI+17} and Baran~\cite{MA20} systems. HoloClean is a holistic rule-based data cleaning system that was first relying on Markov Logic Network~\cite{RCI+17}, which is later replaced by an attention-based learning network~\cite{WZI+20} in its open source distribution~\cite{HoloClean}. Baran~\cite{MA20} is most notable as a configuration-free, human-in-the-loop data cleaning system. It iteratively asks users to manually correct a sampled error and learns to generalize the human correction to the rest of the dataset. Baran does not ask for user-input dependency rules. Instead, it assumes functional dependencies between each pair of attributes. Unfortunately, rule-based data cleaning systems fall short when dealing with spatial data. The main reason is that functional dependencies mainly rely on the exact co-occurrence of record values, which would rarely happen for spatial latitude and longitude location records, captured by inherently inaccurate devices. Our proposed system \sys aims to inject spatial awareness into the core engine of such rule-based data cleaning systems, making them efficiently supporting spatial data.

Earlier data cleaning approaches have considered the use of the approximation in the functional dependency rules~\cite{CIP13,DEE+13,KIJ+15,PSC+15,SZC+15,ZZL+22}. Meanwhile, a spectrum of constraints have also been proposed to support approximate comparison of attribute values~\cite{CDP16}, including matching dependency~\cite{F08}, metric functional dependency~\cite{KSS+09}, differential dependency~\cite{SC11}, and ontology functional dependency~\cite{BKC+17}. Such approximate rules and constraints are mainly designed to tolerate marginal syntactic difference for entities that are actually considered the same, e.g., the words ``Ave.'' and ``Avenue'' should mean the same thing. Unfortunately, such approximate constraints are still not able to capture the needs for spatial attributes. This is mainly for several reasons: (1)~such constraints assume that the difference in values is the exception (e.g., few records use ``Ave.'', but most records use ``Avenue''), while in spatial constraints, it is the norm and expectations that all records are different. (2)~the execution of such constraints cannot scale up to tolerate scalable spatial neighborhood criteria, where distance threshold is much larger and records are actually different form each other. (3)~such constraints are evaluated as a binary logic, where each constraint is either violated or not. Spatial constraints should not follow binary logic as some records may violate a certain constraint in stronger or weaker terms than others, based on the relative distance of records involved in the constraint evaluation.

Within the spatial computing community, spatial data cleaning approaches mainly focus on improving the quality of spatial attributes~\cite{LLJ+23}, but not on following any kind of constraint rules. Examples of such approaches include correcting erroneous GPS point through map matching~\cite{BPS+05, NK09}, employing signal triangulation to enhance the accuracy of indoor locations~\cite{ZGL19, LLC+23}, machine learning approaches for trajectory imputation~\cite{EIM22, MM23}, and data analysis techniques to reduce the uncertainty of location data~\cite{CKP03, ZTP+17}. None of these approaches apply to our case as our goal is not to correct the spatial data itself, but use the spatial information to enhance the accuracy of data cleaning systems for other attributes.



\section{Conclusion}
\label{sec:conclusion}

This paper presented \sys; a novel framework built inside the core engine of rule-based data cleaning systems to boost their accuracy when dealing with spatial data. \sys system architecture is made similar to that of rule-based data cleaning systems where it injects spatial awareness in every system component. In particular, \sys is composed of three main components, {\em spatial error detector}, {\em spatial candidate generator}, and {\em spatial input formulator}. \sys relaxes the functional dependency that drives all data cleaning rules from ``records with the same location would have the same value in a dependent attribute'' to ``records with more similar locations are more likely to have the same value in a dependent attribute'', which is more suitable for spatial attributes. To do so, \sys has to apply two main spatial concepts, {\em spatial neighborhood} and {\em distance weighting}. With spatial neighborhood, records that are within spatial proximity are considered similar and hence are involved in the dependency constraint. With distance weighting, some records satisfy the dependency constraint more than others, and it would weight more for the functional dependency. Experimental results with real datasets for Austin, Chicago, and NYC, and synthetic datasets and a real deployment of \sys inside HoloClean data cleaning system show that \sys significantly boosts the accuracy of its host data cleaning system.




\newpage
\bibliographystyle{plain} 
\bibliography{library}

\end{document}